# Nodal Inertia Pricing for RoCoF Security with Integration of Synchronous and Virtual Resources

Bomiao Liang, *Senior Member, IEEE*, Yu Yang, Licheng Wang, *Senior Member, IEEE*, Lei Wan, Dipti Srinivasan, *Fellow, IEEE*, Fushuan Wen, *Fellow, IEEE*, Youbing Zhang and Beiping Hou

Bomiao Liang (bomiao.liang@zust.edu.cn) and Beiping Hou (bphou@zust.edu.cn) are with the Zhejiang University of Science and Technology, Hangzhou 310023, China.
Yu Yang, Lei Wan and Fushuan Wen (wenfs@hotmail.com) are with the Zhejiang University, Hangzhou 310027, China.
Licheng Wang (corresponding author, l.wang8@uq.edu.au) is with the University of Queensland, Brisbane, QLD 4072, Australia.
Youbing Zhang (youbingzhang@zjut.edu.cn) is with the Zhejiang University of Technology, Hangzhou 310023, China.
Dipti Srinivasan (dipti@nus.edu.sg) is with the National University of Singapore, Singapore 119260.
This work is supported by the National Natural Science Foundation of China under Grants U22B20116.

***Abstract***—Frequency regulation is becoming a more significant challenge in the undergoing low-carbon transition of modern power systems, which is heavily rely on system inertia provision. In the context of market-based energy allocation, reasonable valuation of the system inertia is crucial to its procurement, namely a worthy price should be determined for inertia providers. However, current research on the value of inertia remains insufficient, particularly during the initial phase immediately following a contingency—a critical period when the system primarily relies on the inherent inertia from conventional units and pre-engaged virtual inertia resources. Furthermore, it is found in existing studies that the Center of Inertia (COI) models exhibit significant deviations from actual physical conditions. For market-oriented reforms that explore individual differences, COI models also fail to capture the heterogeneous contributions of inertia at different nodes to system security. Given these circumstances, this paper proposes a nodal inertia pricing mechanism via a day-ahead unit commitment and inertia co-optimization model with nodal resolution. Following the proposed pricing scheme, the day- or hour-ahead inertia provision is monetized and promoted, providing regional guidance to units and devices with virtual inertia provision, which is further verified by simulation results of the South East Australia Power System.

## I. Introduction

The transition towards renewable energy source (RES)-based generation is causing significant inertia shortages in modern power systems, leading to elevated RoCoF and increasing frequency-security risks following disturbances [1, 2]. Since market-based operation is a defining feature of contemporary power systems [3], maintaining an adequate level of inertia in the low-carbon transition requires inertia to be treated not as a costless by-product of energy production [4], but as a service that should be explicitly procured and remunerated according to its security value.

Existing studies have recognized inertia and virtual inertia as explicit market products and have proposed dedicated market design, clearing, and pricing approaches, including inertia-market mechanisms [5], virtual-inertia market design [6], and stochastic inertia pricing in electricity markets [7]. Meanwhile, co-optimization frameworks for energy, inertia, and primary frequency response have shown that inertia should be valued jointly with energy schedules and other ancillary services rather than in isolation [8, 9]. Related studies further suggest that the market value of fast frequency-support services depends not only on procured capacity, but also on dynamic characteristics such as response rate and delivery timing [10-12]. Such findings are particularly significant for distinguishing the security value of inertial and other fast frequency-support services from slower post-contingency governor action.

Building on these insights, recent studies have extended frequency-security market design to converter-dominated power systems, where the rapid response capability of inverter-based resources and virtual inertia becomes increasingly important. In particular, a recent study proposes a multi-timescale joint clearing model for energy, inertia and ramping, while explicitly accounting for the rapid response capability of inverter-based resources in supporting frequency stability and operational flexibility. [13]. Aggregated inverter-based resources and virtual power plants have also been brought into inertia-related ancillary-service markets through aggregation models [14], bilevel clearing formulations [15], and latency-sensitive pricing frameworks [16]. In parallel, cost-allocation studies show that inertia and fast frequency response require not only efficient compensation for providers but also meaningful allocation of procurement costs [17]. Overall, these studies significantly advance the marketization of inertia and frequency-security services and move from simple capacity compensation toward more performance-based settlement. However, most of them still rely on aggregated frequency-security representations and do not yet fully capture locational heterogeneity.

Meanwhile, the physical basis of frequency-security assessment is shifting from aggregated models toward spatially differentiated ones. Existing evidence shows that minimum inertia requirements may vary across areas [18], and that the largest RoCoF can deviate markedly from the center of inertia (COI)-based value because of network topology and inertia distribution [19]. Recent nodal RoCoF-constrained scheduling studies further demonstrate that satisfying a system-average or COI-based criterion does not necessarily guarantee local frequency security [20]. This observation is also consistent with frequency-divider-based analysis of spatial frequency behavior [21] and with ambient-measurement studies questioning whether the COI model can always serve as an adequate proxy for actual system frequency dynamics [22]. Therefore, in low-inertia systems, the security value of inertia is inherently locational: the marginal benefit of additional synchronous or virtual inertia depends on where it is located.

Although the spatial heterogeneity of frequency dynamics has direct implications for market design [23], a dedicated nodal inertia pricing mechanism is still lacking. This gap is particularly important in the very short interval immediately following a disturbance, when governor response remains largely inactive because of deadband and actuation delay, and the initial RoCoF surge can be suppressed only by the inertia of online conventional units and the already deployed virtual inertia. This interval, hereafter referred to as the first stage, is also the period in which the spatial heterogeneity of nodal RoCoFs across the system is most pronounced. It is therefore especially important to distinguish the location-dependent contributions of inertia resources to system security with sufficient granularity. By contrast, once the system enters the second stage, governor action gradually develops, the spatial disparity in RoCoF diminishes, and frequency security becomes more closely tied to primary frequency response, reserves, and other follow-on services. At this stage, because the spatial variation of nodal frequency is less pronounced, the system dynamics can be represented more adequately by aggregated frequency models and have therefore been more extensively studied in existing literature.

To bridge the research gap in the first stage, a nodal inertia pricing scheme for procuring an appropriate level of inertia is proposed and illustrated in this paper. A nodal RoCoF model is deployed instead of COI to reflect the nodal inertia difference in the power system, and a specialized nodal inertia pricing scheme for the first stage frequency security is designed, to exempt from RoCoF limit violations more precisely. Based on the ex-post pricing framework, the solution is achieved by solving the dual problem and utilizing the stationary conditions. The composition and physical significance of the nodal inertia price are further analyzed subsequently. These prices dynamically reflect the state-dependent scarcity and marginal security contribution of inertia at individual nodes, thereby reducing the mismatch between the pricing model and the actual power system, particularly with respect to locational differences. Accordingly, the main contributions of this paper are threefold:

First, a nodal inertia pricing architecture is developed. The traditional COI model is replaced with a node-specific RoCoF model to capture the spatial heterogeneity of inertia requirements in low-inertia systems, thereby effectively identifying and locating “inertia-weak” nodes.

Second, a coordinated clearing mechanism for heterogeneous inertia resources is established. By integrating physical inertia provided by conventional synchronous generators and virtual inertia provided by inverter-based resources into a unified market framework, economically efficient coordination of multiple resource types can be achieved while respecting frequency security constraints.

Third, decoupled and quantifiable price signals are derived. Based on duality theory, the nodal inertia pricing problem is rigorously formulated, which not only reflects the overall system inertia scarcity but also further quantifies the security value of inertia with respect to RoCoF constraints at specific locations. This provides a workable mechanism for the optimal allocation and economic incentives of inertia resources.

## II. Optimal Energy and Inertia Commitment Model

As inferred from [5], virtual inertia is expected to be widely deployed in the future, and the inertia at each bus may originate from traditional sources, virtual sources, or both. According to their characteristics, traditional inertia is determined during the unit commitment process, whereas virtual inertia can be dispatched in response to disturbances within its capacity limits. At the first stage, the suppression of the RoCoF relies solely on online conventional units and the already deployed virtual inertia, whose availability is determined by day-ahead or hour-ahead commitment decisions. In this paper, an energy and inertia commitment model is proposed and optimized to monetize the inertia provision along with normal system operation. Since the greatest RoCoF occurs at the outset of the frequency excursion, when the power imbalance is maximal [1] and the spatial disparity of nodal RoCoFs is significant [19], we derive RoCoF constraints for the instant immediately after a contingency.

### *A. Inertia Characteristics, Provision and Payment*

For nodes with inertia, whose inertia $H_i$ and provision cost might be composed of two parts, namely traditional part $H_{G,i}$ and virtual part $H_{V,i}$.

$$H_i = x_i H_{G,i} + y_i H_{V,i} \tag{1}$$

where $x_i$, $y_i$ are 0-1 variables indicating whether the conventional unit and virtual part at node *i* is on. The value of $H_{G,i}$ is usually constant if the unit is on as depicted in (2), with the inertia provision from conventional units paid according to the load quantity in a vague and preliminary approach.

$$H_{G,i} = \frac{J_i \omega_i^2}{2 S_i} \tag{2}$$

where $J_i$ is the moment of inertia of the rotor for synchronous generator *i*; $\omega_i$ is the rated mechanical angular velocity of the rotor for synchronous generator *i*; $S_i$ is the rated apparent power for synchronous generator *i*.

Unlike the rigid characteristics of synchronous generator inertia, the virtual inertia provided by an energy storage system offers greater regulation flexibility. Its service capability is primarily achieved through power electronic converters operating in virtual synchronous generator (VSG) control mode, which can be rapidly adjusted according to system operational needs. Therefore, the cost of virtual inertia ancillary services is closely related to the dispatch status of the device and the level of capacity reserved. Subsequently, its pricing may be considered based on the actual provided virtual inertia magnitude, $H_{V,i}$.

According to the current policy, the traditional part $H_{G,i}$ is remunerated for inertia provision only when its load rate $\gamma_{G,i}$ is below a certain ratio $\gamma_L$, such as 30%-40%, whereas the virtual part $H_{V,i}$ is dispatchable on demand and is normally remunerated according to its cleared quantity. In this paper, the nodal inertia offer price is denoted by $\lambda_{H,i}$, which applies to both the deployed virtual inertia and the traditional part that is eligible for inertia payment under low-loading conditions, as shown below.

$$\gamma_{G,i} = \frac{p_{G,i}}{p_{G,i}^{max}} \tag{3}$$

$$z_{G,i} = \begin{cases} 1, & \gamma_{G,i} \le \gamma_L \\ 0, & \gamma_{G,i} > \gamma_L \end{cases} \tag{4}$$

where $p_{G,i}$ is the generation output of the conventional units at node *i*, with $p_{G,i}^{max}$ as its capacity; $z_{G,i}$ is 0-1 variables indicating whether conventional units at node *i* are paid for inertia provision under low load rate.

Research on virtual inertia provision and remuneration is still at a preliminary stage. Existing studies rarely consider the spatial disparity of nodal RoCoFs after a contingency and the spatial heterogeneity of inertia providers and thus fail to capture the resulting differences in inertia value and price across buses. To fill this gap, this paper monetizes the contributions of individual participants to RoCoF security in the first stage and derives corresponding locational inertia prices. It should be noted that the inertia payment considered here is only one component of an inertia provider's overall revenue, which should be sufficient to cover provision costs and provide reasonable returns in the long run.

*B. Network Equation*

For a power system comprising $n$ generators (including conventional and virtual units) and $m$ nodes, the post-contingency changes in nodal power injections and voltage angles following the tripping of synchronous generator $k$ can be approximately described by the augmented direct current (DC) power flow equations as follows [21].

$$\begin{bmatrix} \Delta P_G^k(0^+) \\ \Delta P_D^k(0^+) \end{bmatrix} = \begin{bmatrix} B_{\mathrm{GG}} & B_{GB} \\ B_{BG} & B_{BB} \end{bmatrix} \begin{bmatrix} \Delta\Theta_G^k(0^+) \\ \Delta\Theta_D^k(0^+) \end{bmatrix} \tag{5}$$

where $0^+$ denotes the initial instant immediately after a disturbance (i.e., the first stage); the superscript $k$ corresponds to the $k$-th contingency scenario, i.e., the trip of generator $k$; $B_{\mathrm{GG}} \in \mathbb{R}^{n\times n}$, $B_{GB} \in \mathbb{R}^{n\times m}$ and $B_{BG} \in \mathbb{R}^{m\times n}$ are the expanded susceptance matrices that incorporate the generator internal reactance; $B_{BB} \in \mathbb{R}^{m\times m}$ represents the matrix constructed by superimposing the standard network susceptance matrix and a diagonal matrix that contains the generator internal reactance; $\Delta P_G^k \in \mathbb{R}^{n\times 1}$ and $\Delta P_D^k \in \mathbb{R}^{m\times 1}$ are the vectors of generator electromagnetic power variations and load power variations respectively; $\Delta\Theta_G^k \in \mathbb{R}^{n\times 1}$ and $\Delta\Theta_D^k \in \mathbb{R}^{m\times 1}$ denote the vectors of rotor-angle variations at the nodes behind the generators' internal reactance and voltage phase-angle variations at the load nodes respectively. As illustrated in Fig. 1, the nodes behind the internal reactance of generators are defined as generator nodes, whereas the system-side nodes (including the generator terminal node) are defined as load nodes. In addition, inverter saturation is neglected in this paper, and each VSG is assumed to be equipped with sufficient capacity to match its offered virtual inertia level. Therefore, to simplify the model representation, the dynamic differences between synchronous generators and VSGs in frequency response are not explicitly considered, and both are uniformly represented as generator nodes.

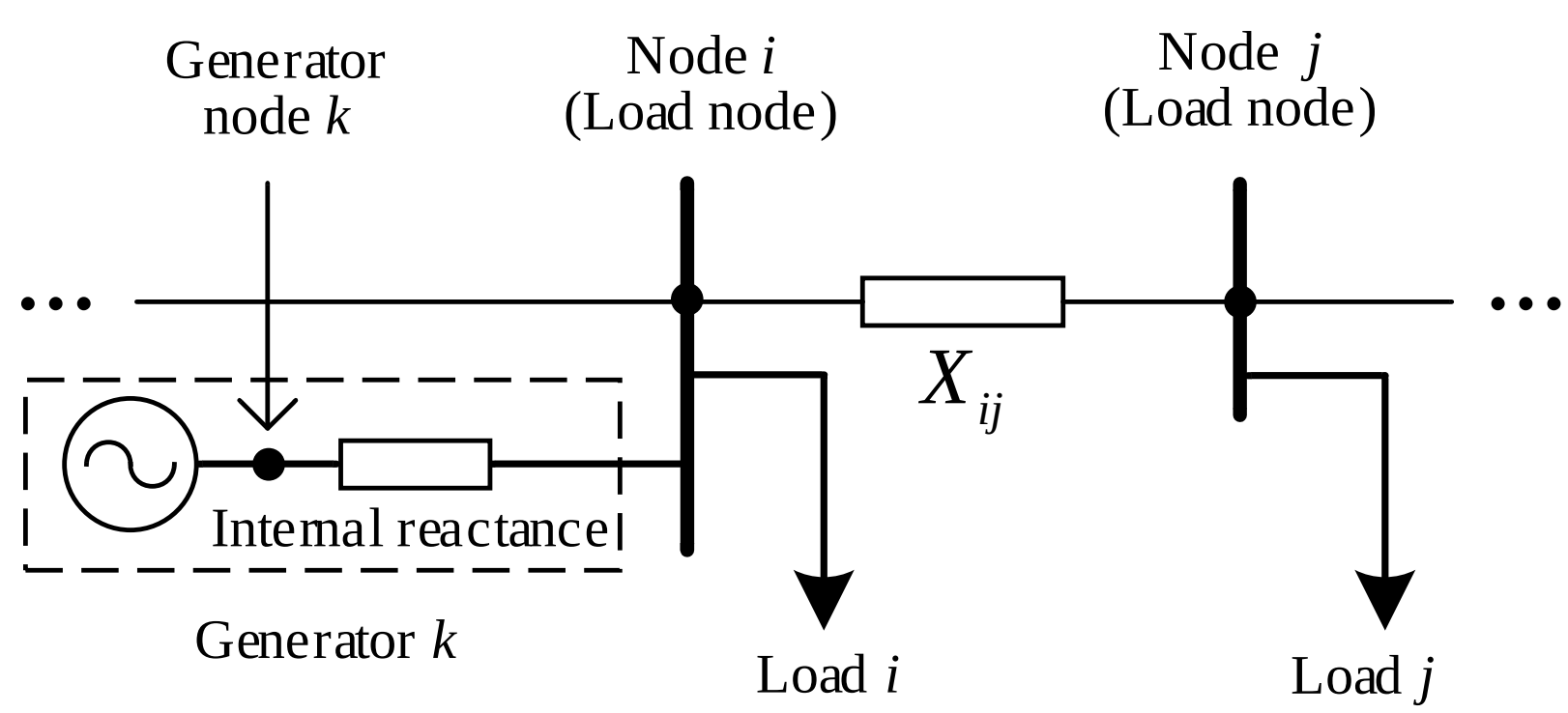

Fig. 1. Schematic definition of load nodes and generator nodes.

Assuming that the network topology remains unchanged during the contingency, (5) can be partitioned and further rewritten as (6).

$$\Delta P_G^k(0^+) = \left(B_{\mathrm{GG}} - B_{GB}\cdot B_{BB}^{-1}\cdot B_{BG}\right)\cdot \Delta\Theta_G^k(0^+) + \; B_{GB}\cdot B_{BB}^{-1}\cdot \Delta P_D^k(0^+) \tag{6}$$

At the instant of fault occurrence ($t=0^+$), the load power can be considered unchanged [24], namely

$$\Delta P_D^k(0^+) = \left[0\ldots,0,\ldots 0\right]^T \tag{7}$$

Substituting (7) into (6) yields

$$\Delta P_G^k\left(0^+\right)=\left(B_{GG}-B_{GB}\cdot B_{BB}^{-1}\cdot B_{BG}\right)\cdot\Delta\Theta_G^k\left(0^+\right) \tag{8}$$

The invertibility of matrix $B_{BB}^{-1}$ has been proved in the authors' previous work [18]. Eqn. (8) establishes a mapping between the active power variations at the generator nodes and the corresponding voltage phase-angle variations immediately after the contingency. To solve the system of equations represented by (8), independent boundary conditions for the generator nodes must be provided.

### C. Post-Contingency Boundary Conditions and Nodal RoCoF

If unit $k$ trips, the active power variation at node $k$ at time instant $0^+$ equals the negative of its pre-contingency active power output.

$$\Delta p_{G,k}^k\left(0^+\right)=-p_{G,k} \tag{9}$$

where $\Delta p_{G,k}^k\left(0^+\right)$ denotes the $k$-th element of the vector $\Delta P_G^k\left(0^+\right)$.

For a non-faulted node $i$, its boundary condition depends on whether inertia support is available at that node. If the node has online inertia support, i.e., an online synchronous generator or an operating virtual inertia device, then according to state continuity and inertial response characteristics, the phase angle at that generator node cannot change instantaneously at $t=0^+$. If the node has neither an online synchronous generator nor virtual inertia support, it can be equivalently treated as a load node, and its active power injection variation is zero. Accordingly, boundary conditions of generator nodes can be uniformly established as

$$\left[1-\left(1-x_i\right)\left(1-y_i\right)\right]\Delta\theta_{G,i}^k\left(0^+\right)=\left(1-x_i\right)\left(1-y_i\right)\Delta p_{G,i}^k\left(0^+\right),\ \ \forall i\neq k \tag{10}$$

where $\Delta\theta_{G,i}^k\left(0^+\right)$ and $\Delta p_{G,i}^k\left(0^+\right)$ denote the $i$-th element of the vectors of $\Delta\Theta_G^k\left(0^+\right)$ and $\Delta P_G^k\left(0^+\right)$ respectively.

Since the maximum power imbalance occurs at the initial instant of the contingency, i.e., $t=0^+$, the maximum RoCoF typically also appears at this moment [15, 19]. For the contingency scenario $k$, the swing equation of conventional or virtual unit $i$ is given as follows:

$$\left|\frac{\Delta p_{G,i}^M\left(0^+\right)-\Delta p_{G,i}^k\left(0^+\right)}{2H_i}\right|\leq RoCoF_{max},\forall i\neq k \tag{11}$$

where $\Delta p_{G,i}^M\left(0^+\right)$ denotes the variation in mechanical power for a synchronous generator or, in the case of a VSG, the variation in its power reference, and its value is equal to zero at time instant $0^+$, i.e., $\Delta p_{G,i}^M\left(0^+\right)=0$. By substituting (1) into (11) and further extending (11) to all units, regardless of whether they are online or offline, we obtain:

$$-2\left(x_iH_{G,i}+y_iH_{V,i}\right)RoCoF_{max}\leq\Delta p_{G,i}^k\left(0^+\right)\leq2\left(x_iH_{G,i}+y_iH_{V,i}\right)RoCoF_{max},\forall i\neq k \tag{12}$$

For an online unit $i$ $\left(i\neq k\right)$, it can be readily verified that (11) is equivalent to (12), whereas for an offline unit $i$, $\Delta p_{G,i}^k\left(0^+\right)=0$, and therefore (12) still holds.

So far, the relationship between nodal power variations and phase-angle variations in (8), together with the corresponding boundary conditions in (9)-(10), has been established. In addition, the coupling between nodal inertia levels and the RoCoF security constraint at $t=0^+$ is described in (12). Together, these equations constitute a complete set of constraints that the system must satisfy.

### D. Optimal Energy and Inertia Commitment Model

To concentrate on nodal inertia pricing in the first stage following a contingency, the time argument $0^+$is omitted hereafter for brevity. Meanwhile, power losses and voltage constraints are neglected so that the analysis can emphasize energy and inertia pricing. As in the conventional derivation of the Locational Marginal Price (LMP), the former can be incorporated through loss coefficients, while the latter can be examined and adjusted separately.

The optimal energy and inertia commitment model with RoCoF constraints can be formulated as the following. The proposed objective function as depicted in (13) comprehensively considers the generation cost of synchronous units, inertia compensation cost under low-load operating conditions, generation cost of RES units, and dispatch cost of virtual inertia.

$$\begin{aligned} min \; & \underbrace{\sum_{i \in V_g} \left( a_{G,i} p_{G,i} + b_{G,i} x_i + z_{G,i} \lambda_{H,i} H_{G,i} \right)}_{\text{generation and inertia compensation cost of synchronous units}} \\ & + \underbrace{\sum_{i \in V_r} \left( c_{RES,i} p_{RES,i} \right)}_{\text{RES generation cost}} + \underbrace{\sum_{i \in V_v} \left( \lambda_{H,i} l_{V,i} \right)}_{\text{dispatch cost of virtual inertia}} \end{aligned} \tag{13}$$

where $V_g$, $V_r$ and $V_v$ are the sets of node with synchronous units, RES units and virtual inertia respectively; $a_{G,i}$ and $b_{G,i}$ denote the generation and no-load cost of the synchronous units at node $i$; $c_{RES,i}$ and $p_{RES,i}$ are the generation cost and output power of renewable energy units. Auxiliary variable $l_{V,i} = y_i H_{V,i}$ is introduced to maintain the linearity of the objective function. $\lambda_{H,i}$ and $z_{G,i}$ have been explained in Section II-B.

The model constraints include power balance and network constraints, equipment output and capacity constraints, synchronous inertia compensation constraints, and nodal RoCoF security constraints. For ease of presentation, the corresponding Lagrange multipliers (i.e., dual variables) are given after each constraint, separated by a colon ":".

1) Power balance and network constraints

$$\sum_{i \in V_g} p_{G,i} + \sum_{i \in V_r} p_{RES,i} = \sum_{i \in V_d} D_i \quad : \lambda_{BAL} \tag{14}$$

$$\sum_i T_{\zeta,i} \left( p_{G,i} + p_{RES,i} - D_i \right) = P_\zeta \quad : \lambda_{INJ,\zeta} \qquad \forall \zeta \in V_l \tag{15}$$

$$-P_l^{max} \le P_\zeta \le P_l^{max} \quad : \lambda_{TRA,\zeta}^{LOW}, \lambda_{TRA,\zeta}^{UPP} \qquad \forall \zeta \in V_l \tag{16}$$

where $V_d$ and $V_l$ are the set of load nodes and transmission lines respectively; $P_\zeta$ represents the transmission power of line $\zeta$; $T_{\zeta,i}$ denotes the Power Transfer Distribution Factor (PTDF) between node $i$ and line $\zeta$; $P_l^{max}$ represents the maximum transmission power of line $\zeta$.

Eqn. (14) is the system power balance constraint, ensuring that the total generation matches the total load demand; Eqn. (15) describes the linear mapping relationship between nodal power injections and line power flows; Eqn. (16) restricts the line power flow from exceeding its transmission capacity limit.

2) Equipment output and capacity constraints

Synchronous units, RESs, and VSG are all subject to limitations imposed by their rated capacities or maximum output capabilities.

$$x_i p_{G,i}^{min} \le p_{G,i} \le x_i p_{G,i}^{max} \quad : \lambda_{GMIN,i}^{LOW}, \lambda_{GMAX,i}^{UPP} \qquad \forall i \in V_g \tag{17}$$

$$0 \le p_{RES,i} \le p_{RES,i}^{max} \quad : \lambda_{RMIN,i}^{LOW}, \lambda_{RMAX,i}^{UPP} \qquad \forall i \in V_r \tag{18}$$

$$y_i H_{V,i}^{min} \le H_{V,i} \le y_i H_{V,i}^{max} \qquad \forall i \in V_v \tag{19}$$

where $p_{G,i}^{min}$ and $p_{G,i}^{max}$ are minimum and maximum output of unit $i$; $p_{RES,i}^{max}$ represents the maximum output of RES $i$; $H_{V,i}^{min}$ and $H_{V,i}^{max}$ are the lower and upper limit of the virtual inertia $i$. The corresponding Lagrange multiplier of (19) is not listed here as it will later be transformed into other linear constraint forms through auxiliary variables as mentioned.

3) Synchronous inertia compensation constraints

Synchronous inertia compensation constraints are described in (3)-(4). Constraints in (4) can be converted into the following linear form using the Big M method, as in (20). $M$ is a sufficiently large positive constant, whereas $\epsilon$ is a sufficiently small positive number.

$$\left(\gamma_L+\epsilon\right)x_i-M\,z_{G,i}\le\frac{p_{G,i}}{p_{G,i}^{max}}\le\gamma_L x_i+M\left(1-z_{G,i}\right)\quad :\lambda_{COM,i}^{LOW},\lambda_{COM,i}^{UPP}\qquad \forall i\in V_g \tag{20}$$

State-coupling constraints are also introduced to prevent units in shutdown state from being incorrectly regarded as eligible for compensation merely because they satisfy the low-loading condition, as described in (21).

$$z_{G,i}\le x_i\qquad \forall i\in V_g \tag{21}$$

Constraint (21) ensures that the inertia compensation state variable $z_{G,i}$ can only take the value of 1 when the unit is online. Since this constraint involves only integer variables and does not directly generate marginal price signals in the ex-post pricing phase, its corresponding Lagrange multipliers are not listed.

4) Nodal RoCoF security constraints

According to the derivation in previous sections, Eqn. (8)-(10), (12) jointly constitute the nodal RoCoF frequency security constraint set. Denote $B^{RED}\coloneqq B_{\mathrm{GG}}-B_{GB}\cdot B_{BB}^{-1}\cdot B_{BG}$, where $B_{ij}^{RED}$ is its $\left(i,j\right)$-th entry. Accordingly, equations (8) and (9) can be rewritten as (22) and (23) after rearrangement.

$$\Delta p_{G,i}^{k}\left(0^{+}\right)-\sum_{j\in V_g\bigcup V_v}B_{ij}^{RED}\,\Delta\theta_{G,j}^{k}\left(0^{+}\right)=0\quad :\lambda_{NOD,i}^{k}\qquad \forall i\in V_g\bigcup V_v \tag{22}$$

$$\Delta p_{G,k}^{k}\left(0^{+}\right)+p_{G,k}=0\quad :\lambda_{FAU}^{k}\qquad \forall k\in V_g \tag{23}$$

Following the same treatment as in the objective function, the auxiliary variables $l_{V,i}$ are also used here. Specifically, the following linear constraint set defined in (24)-(26) is constructed using the Big M method to linearize the constraints in (12).

$$y_i H_{V,i}^{min}\le l_{V,i}\le y_i H_{V,i}^{max}\quad :\lambda_{AUX,i}^{LOW},\lambda_{AUX,i}^{UPP}\qquad \forall i\in V_v \tag{24}$$

$$H_{V,i}-H_{V,i}^{max}\left(1-y_i\right)\le l_{V,i}\le H_{V,i}-H_{V,i}^{min}\left(1-y_i\right)\quad :\lambda_{CAP,i}^{LOW},\lambda_{CAP,i}^{UPP}\qquad \forall i\in V_v \tag{25}$$

$$\begin{aligned}&-2\left(x_i H_{G,i}+l_{V,i}\right)RoCoF_{max}\le\Delta p_{G,i}^{k}\left(0^{+}\right)\le 2\left(x_i H_{G,i}+l_{V,i}\right)RoCoF_{max}\\ &:\lambda_{ROC,i}^{LOW,k},\lambda_{ROC,i}^{UPP,k}\qquad \forall i\in V_g\bigcup V_v, i\ne k\end{aligned} \tag{26}$$

As for the nonlinear part in (10), the constraint essentially corresponds to two cases: if both $x_i$ and $y_i$ are 0, then $\Delta p_{G,i}^{k}\left(0^{+}\right)=0$; otherwise, if either $x_i$ or $y_i$ is 1, then $\Delta\theta_{G,i}^{k}\left(0^{+}\right)=0$. Among these, the first case is automatically covered by (26), with only the second case requires a linear representation, as expressed in (27)-(28).

$$-M\left(1-x_i\right)\le\Delta\theta_{G,i}^{k}\left(0^{+}\right)\le M\left(1-x_i\right)\quad :\lambda_{G,i}^{LOW,k},\lambda_{G,i}^{UPP,k}\qquad \forall i\in V_g, i\ne k \tag{27}$$

$$-M\left(1-y_i\right)\le\Delta\theta_{G,i}^{k}\left(0^{+}\right)\le M\left(1-y_i\right)\quad :\lambda_{VI,i}^{LOW,k},\lambda_{VI,i}^{UPP,k}\qquad \forall i\in V_v, i\ne k \tag{28}$$

In summary, the proposed joint model consists of the objective function (13) and the constraints (14)-(18) and (20)-(28). This model is a typical mixed integer linear programming (MILP) problem and can be efficiently solved by commercial solvers such as Cplex or Gurobi.

## III. DECOMPOSITION OF NODAL INERTIA PRICING

### *A. Linearized Dual Model and Market Equilibrium Conditions*

In the joint model constructed in Section II-D, the inclusion of discrete integer variables renders the original problem nonconvex. For such a mixed-integer linear programming (MILP) problem, the strong duality theorem generally no longer holds directly. Therefore, it is not possible to obtain marginal price signals with strict economic meaning directly from the dual solution of the original problem.

The ex-post pricing method is adopted to address such non-convex pricing problems. This approach adopts a two-stage solution strategy: first, solve the original MILP problem to obtain the optimal dispatch plan that satisfies all physical constraints; then, fix all discrete variables to convert the original problem into a continuous linear programming (LP) problem, and obtain the dual solution based on this linear model. The ex-post pricing method can avoid the dimensionality curse associated with the convex hull relaxation approach and the loss of physical meaning caused by continuous relaxation, making it most suitable for this study. Specifically, in the ex-post pricing process, the optimal integer solutions obtained from the original problem, namely, $x_i$, $y_i$, and $z_{G,i}$, are fixed as parameters $x_i^*$, $y_i^*$, and $z_{G,i}^*$ respectively, which ensures that the price signals are strictly based on a physically feasible solution that satisfies all node RoCoF security constraints. This treatment also transforms the original mixed-integer problem into a continuous LP problem that satisfies the strong duality theorem, thereby providing a rigorous mathematical foundation for the subsequent calculation of locational marginal prices.

To obtain the locational marginal price, the Lagrangian function $L$ incorporating all constraints for the linearized model is constructed. According to the stationarity condition in the Karush-Kuhn-Tucker (KKT) optimality conditions, taking the partial derivatives of the Lagrangian function with respect to each continuous decision variable and setting them to zero yields the equilibrium conditions for market clearing. The corresponding stationarity conditions for each variable, namely $p_{G,i}$, $p_{RES,i}$, $P_\zeta$, $H_{V,i}$, $l_{V,i}$, $\Delta p_{G,k}^k$, $\Delta p_{G,i}^k$, $\Delta\theta_{G,k}^k$, and $\Delta\theta_{G,i}^k$, are listed sequentially in (29)-(37).

$$\begin{aligned}\frac{\partial L}{\partial p_{G,i}} &= a_{G,i} + \lambda_{BAL} + \sum_{\zeta} T_{\zeta,i}\lambda_{INJ,\zeta} + \lambda_{FAU}^{i} \\ &- \frac{\lambda_{COM,i}^{LOW}}{p_{G,i}^{max}} + \frac{\lambda_{COM,i}^{UPP}}{p_{G,i}^{max}} - \lambda_{GMIN,i}^{LOW} + \lambda_{GMAX,i}^{UPP} = 0\end{aligned} \tag{29}$$

$$\frac{\partial L}{\partial p_i^{RES}} = c_{RES,i} + \lambda_{BAL} + \sum_{\zeta} T_{\zeta,i}\lambda_{INJ,\zeta} - \lambda_{RMIN,i}^{LOW} + \lambda_{RMAX,i}^{UPP} = 0 \tag{30}$$

$$\frac{\partial L}{\partial P_\zeta} = -\lambda_{INJ,\zeta} - \lambda_{TRA,\zeta}^{LOW} + \lambda_{TRA,\zeta}^{UPP} = 0 \tag{31}$$

$$\frac{\partial L}{\partial H_{V,i}} = \lambda_{CAP,i}^{LOW} - \lambda_{CAP,i}^{UP} = 0 \tag{32}$$

$$\begin{aligned}\frac{\partial L}{\partial l_{V,i}} &= \lambda_{H,i} - 2\,RoCoF_{max}\sum_{k\neq i}\left(\lambda_{ROC,i}^{LOW,k} + \lambda_{ROC,i}^{UPP,k}\right) \\ &- \lambda_{AUX,i}^{LOW} + \lambda_{AUX,i}^{UPP} - \lambda_{CAP,i}^{LOW} + \lambda_{CAP,i}^{UPP} = 0\end{aligned} \tag{33}$$

$$\frac{\partial L}{\partial \Delta p_{G,k}^k(0^+)} = \lambda_{NOD,k}^k + \lambda_{FAU}^k = 0 \tag{34}$$

$$\frac{\partial L}{\partial \Delta p_{G,i}^k(0^+)} = \lambda_{NOD,i}^k - \lambda_{ROC,i}^{LOW,k} + \lambda_{ROC,i}^{UPP,k} = 0,\ \forall i\neq k \tag{35}$$

$$\frac{\partial L}{\partial \Delta\theta_{G,k}^k(0^+)} = -\sum_j B_{jk}^{RED}\lambda_{NOD,j}^k = 0 \tag{36}$$

$$\begin{aligned}\frac{\partial L}{\partial \Delta\theta_{G,i}^k(0^+)} &= -\sum_j B_{ji}^{RED}\lambda_{NOD,j}^k - \lambda_{G,i}^{LOW,k} \\ &+ \lambda_{G,i}^{UPP,k} - \lambda_{VI,i}^{LOW,k} + \lambda_{VI,i}^{UPP,k} = 0,\ \forall i\neq k\end{aligned} \tag{37}$$

Based on the aforementioned KKT conditions, the corresponding dual problem which essentially reformulates the acceptable cost boundary of the system under the constraint of nodal frequency security, can be expressed as in (38).

$$\max\left\{\begin{array}{l} -\lambda_{BAL}\sum_{i} D_i - \sum_{\zeta}\left(\lambda_{INJ,\zeta}\sum_{i} T_{\zeta,i} D_i\right) \\ -\sum_{\zeta}\lambda_{TRA,\zeta}^{LOW} P_l^{max} - \sum_{\zeta}\lambda_{TRA,\zeta}^{UPP} P_l^{max} \\ +\sum_{k}\left[\begin{array}{l} \sum_{i\neq k} -M\left(1-x_i^*\right)\lambda_{G,i}^{LOW,k} - \sum_{i\neq k} M\left(1-x_i^*\right)\lambda_{G,i}^{UPP,k} \\ -\sum_{i\neq k} M\left(1-y_i^*\right)\lambda_{VI,i}^{LOW,k} - \sum_{i\neq k} M\left(1-y_i^*\right)\lambda_{VI,i}^{UPP,k} \\ -2\,RoCoF_{max}\sum_{i\neq k} x_i^* H_{G,i}\lambda_{ROC,i}^{LOW,k} - 2\,RoCoF_{max}\sum_{i\neq k} x_i^* H_{G,i}\lambda_{ROC,i}^{UPP,k} \end{array}\right] \\ +\sum_{i} y_i^* H_{V,i}^{min}\lambda_{AUX,i}^{LOW} - \sum_{i} y_i^* H_{V,i}^{max}\lambda_{AUX,i}^{UPP} \\ -\sum_{i} H_{V,i}^{max}\left(1-y_i^*\right)\lambda_{CAP,i}^{LOW} + \sum_{i} H_{V,i}^{min}\left(1-y_i^*\right)\lambda_{CAP,i}^{UPP} \\ +\sum_{i}\left(\gamma_L x_i^* + \epsilon x_i^* - M z_{G,i}^*\right)\lambda_{COM,i}^{LOW} - \sum_{i}\left(\gamma_L x_i^* + M - M z_{G,i}^*\right)\lambda_{COM,i}^{UPP} \\ +\sum_{i} x_i^* p_{G,i}^{min}\lambda_{GMIN,i}^{LOW} - \sum_{i} x_i^* p_{G,i}^{max}\lambda_{GMAX,i}^{UPP} \\ -\sum_{i} p_{RES,i}^{max}\lambda_{RMAX,i}^{UPP} \end{array}\right\} \tag{38}$$

The constraints of the dual problem are precisely the stationarity conditions mentioned above, together with the non-negativity conditions imposed on the dual variables associated with all inequality constraints.

### *B. Decomposition and Implications of Nodal Inertia Pricing*

By solving the aforementioned dual problem, the shadow prices corresponding to various system constraints can be obtained, thereby quantitatively characterizing the marginal scarcity of inertia resources under the condition of satisfying nodal RoCoF frequency security constraints. Based on the KKT stationarity conditions provided in the previous section, the dual variables can be further translated into price signals with clear economic interpretations, comprising both the conventional locational marginal prices and the nodal inertia prices proposed in this paper. The former, denoted as $C_{Energy,i}$, represents the marginal system cost incurred by increasing a unit load at node $i$, as conventionally understood, whereas the latter, $C_{Inertia,i}$, reflects the marginal cost incurred by the system to maintain the RoCoF security constraint.

$$C_{Energy,i} = -\lambda_{BAL} - \sum_{\zeta} T_{\zeta,i}\lambda_{INJ,\zeta} \tag{39}$$

$$C_{Inertia,i} = 2\,RoCoF_{max}\sum_{k\neq i}\left(\lambda_{ROC,i}^{LOW,k} + \lambda_{ROC,i}^{UPP,k}\right) \tag{40}$$

In scenarios with high penetration of wind, photovoltaic, and other low-marginal-cost generation, renewable energy is typically prioritized in market clearing, causing nodal electricity prices to decline and approach the marginal generation cost of renewable energy units. Under these conditions, revenue from the energy market alone is often insufficient to cover the no-load cost of keeping synchronous generators online, as well as the opportunity cost of providing inertia support. This highlights the need for the proposed inertia ancillary service pricing mechanism.

Nodal inertia price is directly determined by the Lagrange multipliers $\lambda_{ROC,i}^{LOW,k}$ and $\lambda_{ROC,i}^{UPP,k}$ corresponding to the nodal RoCoF constraints. According to the complementary slackness condition, the relevant multipliers can only be positive when the corresponding RoCoF constraint is binding. Thus, node inertia resources only demonstrate significant market value when there is actual frequency security risk. This indicates that the nodal inertia price can dynamically capture the inertia shortage risk in specific periods and regions of the system. The summation term in (39) also indicates that the inertia price at node $i$ depends on its comprehensive contribution to the frequency security constraints across all potential contingency scenarios. If a node is located in an electrically weak area of the grid and its inertia resources can effectively mitigate the risk of RoCoF limit violations in multiple contingency scenarios, its nodal inertia price will be significantly higher than that of nodes in electrically strong central areas, where inertia support is relatively sufficient.

## IV. Case Study

### *A. System and Simulation Settings*

This section adopts the power system with topology as shown in Fig. 2 as the test case for simulation and validation [14]. The test system comprises 40 traditional synchronous generators, 12 renewable energy source (RES) plants, and 15 virtual inertia resources (VIRs). For the frequency stability verification in the case studies of this section, an N-1 generator trip contingency is uniformly selected as the test set, and the security critical threshold for the system's maximum RoCoF after the fault is set at -1 Hz/s. The value of this security boundary serves only as a baseline setting to verify the model’s feasibility and does not affect the theoretical generality of the proposed nodal pricing mechanism and dual derivation framework.

To verify the effectiveness of the proposed joint energy and inertia clearing model in identifying inertia-weak areas and quantifying the locational value of resources, this section conducts a comparative analysis between the traditional system-level inertia clearing method based on the aggregated COI equivalent model (later referred to as the COI method) and the proposed nodal energy and inertia joint clearing method. To comprehensively evaluate the system frequency response characteristics and the evolution of price signals under different inertia adequacy levels, the following two typical operating scenarios are constructed:

**Scenario 1 (High-inertia Scenario)**: All 40 traditional synchronous generators in the system are online, providing ample system-wide synchronous inertia at 53.46 GWs. The total system load demand is 14.84 GW, with zero output from renewable energy sources integrated into the grid (RES penetration rate is 0%).

**Scenario 2 (Low-inertia Scenario)**: The total system load level remains consistent with Scenario 1. The 26 synchronous generators in Scenario 1 that provided 10.03 GW of base active power output have been equivalently replaced by renewable energy units. This causes the system's RES penetration rate to climb to 67.6%, and the total available synchronous inertia to decay to 18.06 GWs.

For traditional synchronous generating units, the market offer price for their inherent synchronous inertia is uniformly set at 1.2 $/MWs. To accurately reflect the foregone energy revenue resulting from a unit providing inertia support, the model introduces a state-dependent compensation threshold mechanism based on opportunity cost. Specifically, $\gamma_L$ is set to be 40%.

For the 15 VIRs configured within the system, their specific offer prices are set sequentially as: 1.35, 1.83, 1.59, 1.55, 1.92, 1.29, 1.76, 1.75, 1.53, 1.78, 1.38, 1.57, 1.08, 1.05, and 1.93 $/MWs. This differentiated price sequence is directly incorporated as parameters into the objective function optimization of the joint clearing model, imposing economic boundary constraints on the formation of the final nodal inertia price.

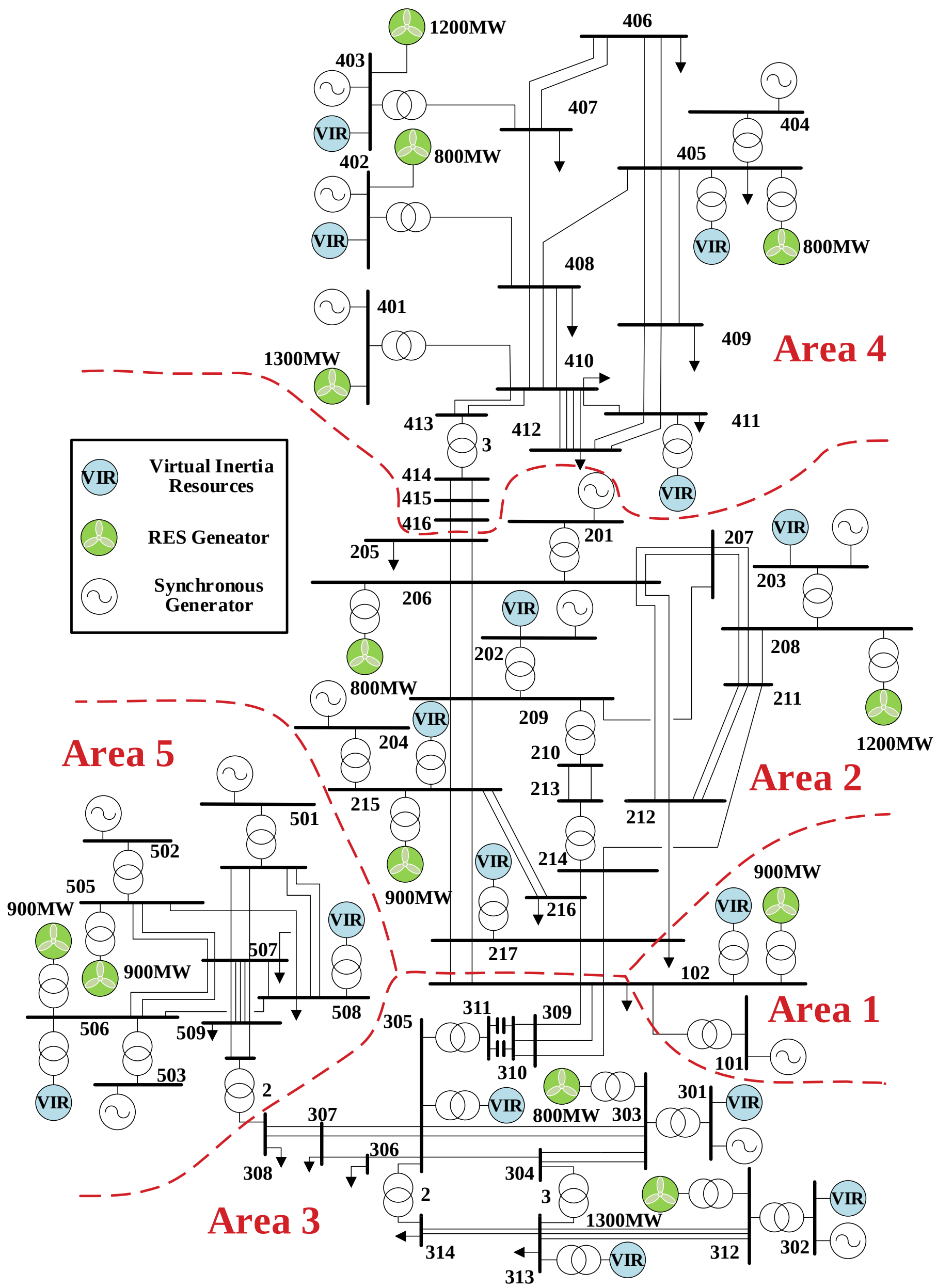

Fig. 2. Test case power system topology.

## B. Results and Analysis

### 1) High-Inertia Scenario: Proposed vs. COI Method

The operational status of the generating units under the high-inertia scenario is shown in Table I. Under the scheduling result of the COI method, the system's COI-RoCoF is -0.3183 Hz/s immediately after a fault, which remains within the security threshold. However, the maximum nodal RoCoF under the most severe fault has reached -1.1411 Hz/s, exceeding the security limit. This phenomenon reveals the limitation of traditional aggregated metrics in masking the uneven spatial distribution of inertia, indicating that a sufficient total system-level inertia does not guarantee absolute frequency security at every node. Constrained by the network susceptance matrix, the supporting power during the initial stage of a sudden disturbance cannot be transmitted instantaneously across the entire grid. In contrast, the proposed method optimizes the spatial distribution of system inertia by starting an additional unit, G203, thereby increasing its inertia provision from 2600 MWs to 3900 MWs and raising the total system inertia from 49,040 MWs to 50,340 MWs, while ensuring that the maximum nodal RoCoF remains within the security limit of -1 Hz/s. Moreover, the dispatch results in Table I show that under both scheduling strategies, generating units tend to maintain higher active power output to gain revenue in the energy market, rather than operating at low output levels to obtain inertia compensation. This scheduling outcome objectively ensures fair generation opportunities for all units in the system.

TABLE I

COMPARISON OF UNIT OPERATING STATUS UNDER HIGH INERTIA SCENARIOS

| Unit No. | Number of units | Maximum output power per unit /MW | COI method | | | | Proposed method | | | |
|---|---|---|---|---|---|---|---|---|---|---|
| | | | Total active power Output (Percentage)/MW | Total inertia /MWs | Inertia price ($/MWs) | Maximum RoCoF after fault | Total active power Output (Percentage)/MW | Total inertia/MWs | Inertia price ($/MWs) | Maximum RoCoF after fault |
| G101 | 2 | 300 | 600(100%) | 2160 | 0 | −1.09 | 600(100%) | 2160 | 7.9363 | −1.00 |
| G201 | 3 | 600 | 1800(100%) | 5760 | 0 | −0.20 | 1800(100%) | 5760 | 0 | −0.19 |
| G202 | 3 | 500 | 1500(100%) | 4200 | 0 | −0.37 | 1500(100%) | 4200 | 0 | −0.36 |
| G203 | 3 | 500 | 540(54%) | 2600 | 0 | −0.46 | 662(44%) | 3900 | 0 | −0.44 |
| G204 | 3 | 600 | 1800(100%) | 5760 | 0 | −0.20 | 1714(95%) | 5760 | 0 | −0.18 |
| G301 | 6 | 600 | 3600(100%) | 10080 | 0 | −0.20 | 3600(100%) | 10080 | 0 | −0.20 |
| G302 | 2 | 400 | 800(100%) | 2800 | 0 | −0.75 | 800(100%) | 2800 | 0 | −0.74 |
| G401 | 3 | 400 | 0(0%) | 0 | 0 | −0.00 | 0(0%) | 0 | 0.4059 | −0.00 |
| G402 | 3 | 300 | 900(100%) | 2700 | 0 | −0.92 | 900(100%) | 2700 | 0 | −0.91 |
| G403 | 2 | 400 | 800(100%) | 2080 | 0 | −0.85 | 800(100%) | 2080 | 0 | −0.85 |
| G404 | 3 | 300 | 900(100%) | 3600 | 0 | −0.57 | 900(100%) | 3600 | 0 | −0.57 |
| G501 | 1 | 300 | 300(100%) | 1050 | 0 | −0.33 | 264(88%) | 1050 | 0 | −0.33 |
| G502 | 4 | 250 | 1000(100%) | 4000 | 0 | −0.54 | 1000(100%) | 4000 | 0 | −0.48 |
| G503 | 2 | 150 | 300(100%) | 2250 | 0 | −1.14 | 300(100%) | 2250 | 1.1718 | −1.00 |
| Total | 40 | / | 14840 | 49040 | / | / | 14840 | 50340 | / | / |

Regarding the clearing price formation mechanism, as shown in Fig. 3, due to the overall sufficiency of system synchronous inertia in the high-inertia scenario, the uniform inertia clearing price calculated by the COI method is $0/MWs. The proposed method takes into account nodal RoCoF security constraints, and the calculated non-zero shadow prices accurately reflect the relative scarcity of local inertia constrained by electrical distance. However, this nodal inertia clearing price remains lower than the bidding prices of all virtual inertia resources. Consequently, neither method actually dispatches virtual inertia. This is because, when the overall synchronous inertia in the system is sufficient, the optimization model can satisfy all nodal frequency security requirements merely by fine-tuning the commitment of conventional synchronous units (such as starting an additional G203 unit). The marginal opportunity cost generated by this internal adjustment of system units has not yet reached the threshold required to trigger the participation of external, higher-priced inertia resources.

As shown in Table II, the operating cost analysis indicates that although neither method dispatches external VIR resources in the aforementioned high-inertia scenario, the proposed method incurs a relatively small additional cost, increasing the total operating cost by approximately 0.23% (from $603,100 to $604,510). This slight cost increment essentially represents the reasonable marginal cost paid to maintain the frequency stability of all system nodes. It thus effectively eliminates the risk of frequency limit violations and ensures the frequency security of all nodes across the entire system.

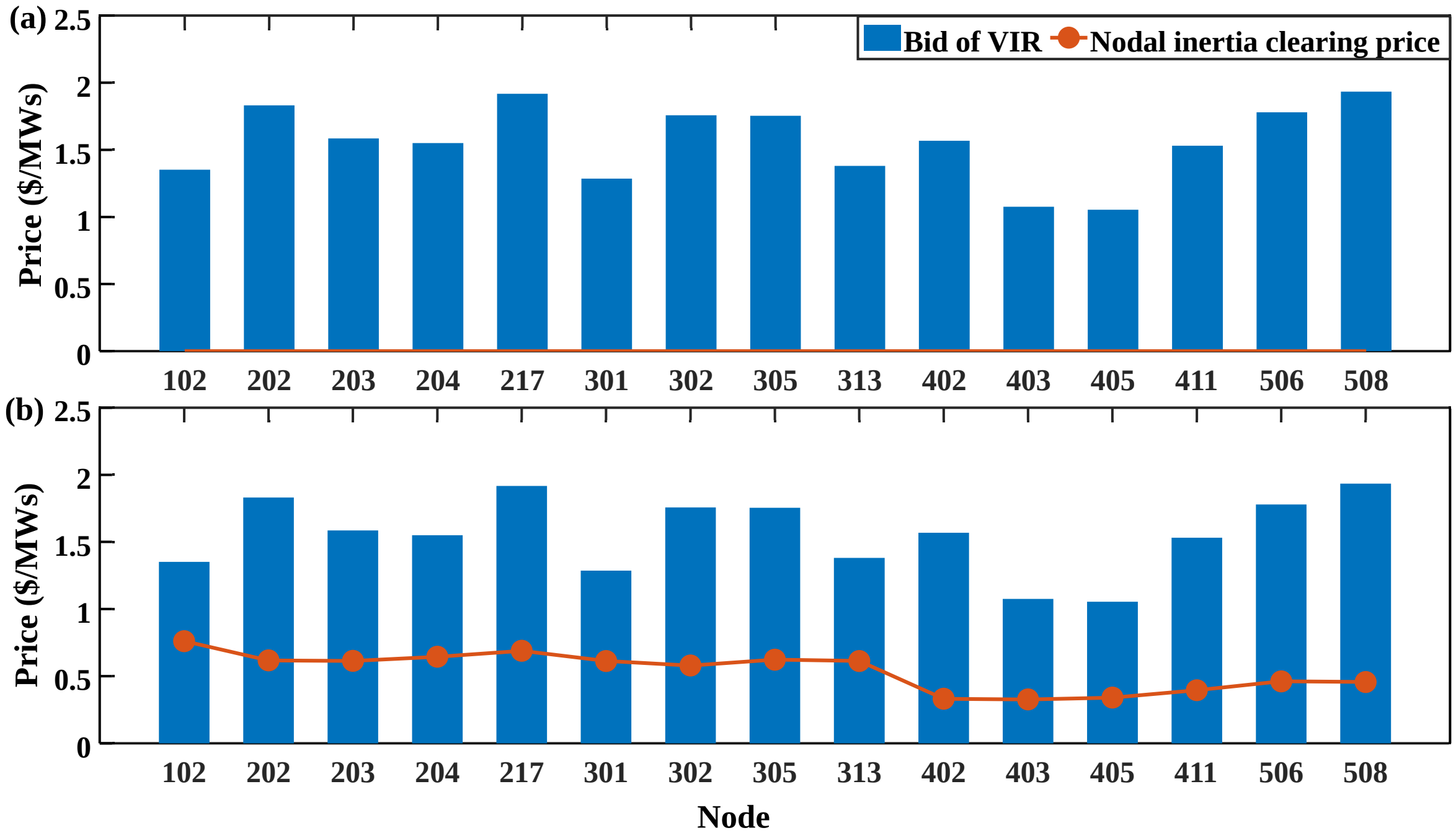

Fig. 3. Nodal inertia clearing prices and bids of VIR under high inertia scenarios (a) COI-based method, (b) Proposed method.

TABLE II

COMPARISON OF ROCOF AND COSTS UNDER HIGH INERTIA SCENARIOS

| Method | COI-RoCoF under most severe fault (Hz/s) | Maximum nodal RoCoF under most severe fault (Hz/s) | Generation cost ($) | Inertia dispatch cost ($) | Total cost ($) | Total inertia revenue ($) |
|---|---|---|---|---|---|---|
| COI | 0.3183 | 1.1411 | 603100 | 0 | 603100 | 0 |
| Proposed | 0.3098 | 1.0000 | 604510 | 0 | 604510 | 0 |

2) Low-Inertia Scenario: Proposed vs. COI method

In low-inertia scenarios, the high penetration of renewable energy further accentuates the imbalance in the spatial distribution of system inertia. As shown in Table III, the COI method considers only the system-level COI-RoCoF as a single security constraint. In pursuit of minimized economic cost, the model dispatches only a small amount of inertia resources, keeping the COI-RoCoF just at the security threshold of -1 Hz/s. However, under the most severe fault, the maximum nodal RoCoF in the system reaches -2.35 Hz/s, indicating a serious limit violation. In contrast, the proposed method effectively identifies weak areas in inertia distribution, dispatches more synchronous inertia and virtual inertia compared to the COI method, and increases the total system inertia from 15,715 MWs to 18,055 MWs. This successfully constrains the maximum RoCoF of all nodes within the secure range.

TABLE III

COMPARISON OF UNIT OPERATING STATUS UNDER LOW INERTIA SCENARIOS

| Unit No. | Number of units | Maximum output power per unit /MW | COI method | | | | Proposed method | | | |
|---|---|---|---|---|---|---|---|---|---|---|
| | | | Total active power Output (Percentage)/MW | Total inertia /MWs | Inertia price ($/MWs) | Maximum RoCoF after fault | Total active power Output (Percentage)/MW | Total inertia/MWs | Inertia price ($/MWs) | Maximum RoCoF after fault |
| G101 | 1 | 300 | 300(100%) | 1080 | 1.054 | -2.15 | 300(100%) | 1080 | 3.6384 | -0.80 |
| G201 | 1 | 600 | 590(98%) | 1920 | 1.054 | -0.30 | 458(76%) | 1920 | 3.9259 | -0.12 |
| G202 | 1 | 500 | 500(100%) | 1400 | 1.054 | -0.71 | 436(87%) | 1400 | 3.9788 | -0.26 |
| G203 | 1 | 500 | 0(0%) | 0 | 1.054 | -0.00 | 305(61%) | 1300 | 4.0773 | -0.30 |
| G204 | 1 | 600 | 450(75%) | 1920 | 1.054 | -0.34 | 329(55%) | 1920 | 3.8258 | -0.12 |
| G301 | 1 | 600 | 600(100%) | 1680 | 1.054 | -0.27 | 497(83%) | 1680 | 3.1808 | -0.15 |
| G302 | 1 | 400 | 400(100%) | 1400 | 1.054 | -1.46 | 400(100%) | 1400 | 3.1190 | -0.69 |
| G401 | 1 | 400 | 0(0%) | 0 | 1.054 | -0.00 | 265(66%) | 1040 | 3.7230 | -0.68 |
| G402 | 1 | 300 | 300(100%) | 900 | 1.054 | -1.31 | 300(100%) | 900 | 10.1117 | -1.00 |
| G403 | 1 | 400 | 400(100%) | 1040 | 1.054 | -0.85 | 303(76%) | 1040 | 3.7013 | -0.80 |
| G404 | 1 | 300 | 300(100%) | 1200 | 1.054 | -0.79 | 300(100%) | 1200 | 3.6996 | -0.62 |
| G501 | 1 | 300 | 300(100%) | 1050 | 1.054 | -0.80 | 247(82%) | 1050 | 0 | -0.33 |
| G502 | 1 | 250 | 250(100%) | 1000 | 1.054 | -1.11 | 250(100%) | 1000 | 2.0902 | -0.44 |
| G503 | 1 | 150 | 150(100%) | 1125 | 1.054 | -2.35 | 150(100%) | 1125 | 2.0715 | -0.95 |
| Total | 14 | / | 4540 | 15715 | / | / | 4540 | 18055 | / | / |

The differences in scheduling strategies further drive a divergence in their market pricing mechanisms and resource allocation outcomes. As shown in Fig. 4, the COI method, facing a single global inertia constraint, only needs to dispatch the cheapest marginal resources (i.e., the VIR at node 405 with a bid of 1.05 /MWs) in ascending order of their bids. When the global average meets the mathematical constraint, the bid of this marginal unit determines the uniform market clearing price across the entire grid. Therefore, the uniform inertia clearing price calculated by the COI method is only 1.05 /MWs. In the proposed method, however, the clearing price is dominated by resource scarcity. When the RoCoF at a specific weak node approaches the security limit, the corresponding dual shadow price of that constraint increases nonlinearly. As shown in Fig. 4(b), the extremely high clearing price at node 203 (15.57 $/MWs) indicates that injecting a unit of inertia at this location yields the highest marginal contribution to alleviating nodal frequency security risks. Such price signals quantify the spatial value of inertia resources, thereby effectively guiding the system to reasonably and adequately dispatch a total of six VIRs at nodes 102, 203, 204, 217, 506, and 508.

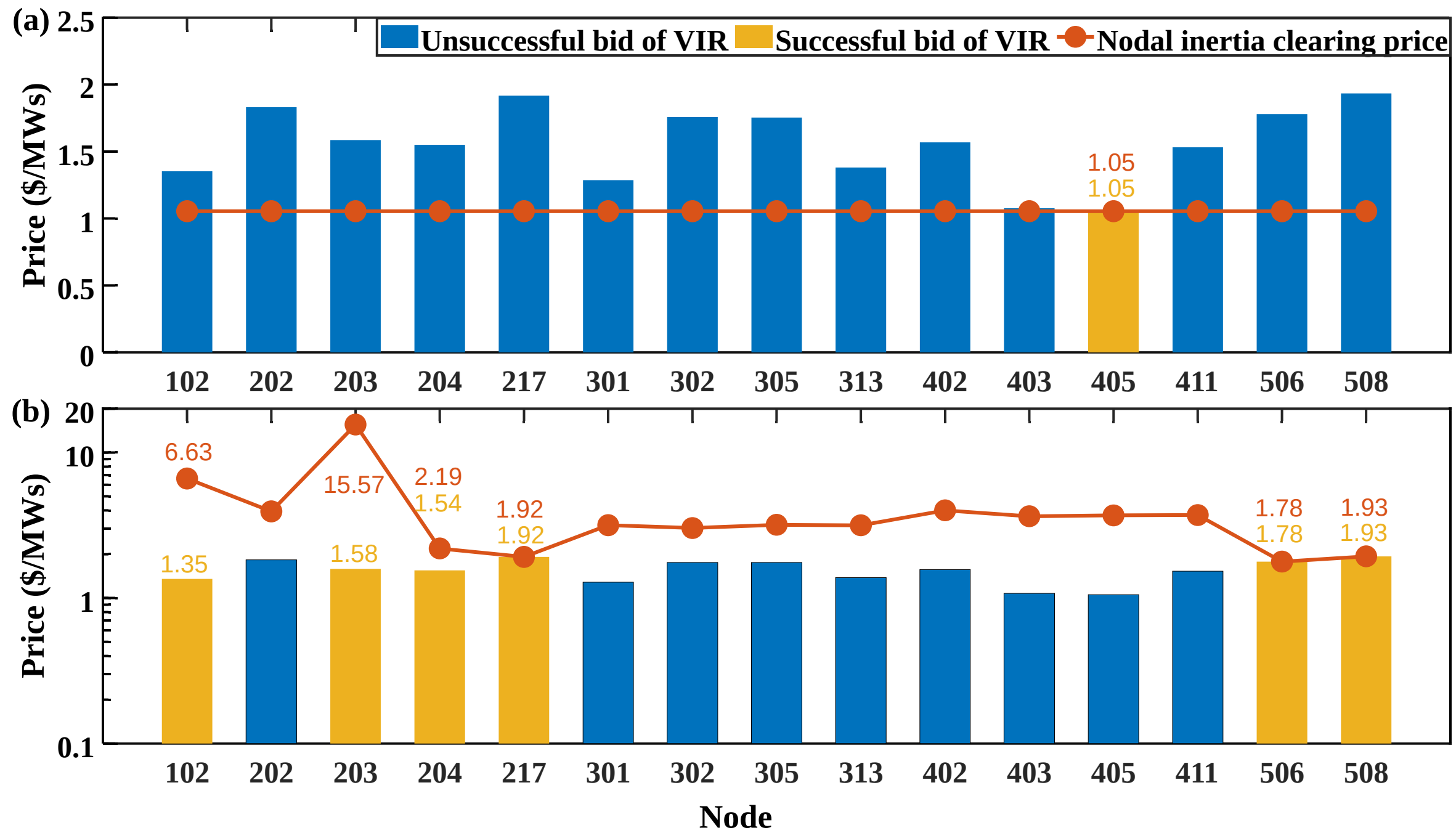

Fig. 4. Nodal inertia clearing prices and bids of VIR under low inertia scenarios (a) COI-based method, (b) Proposed method.

As shown in Table IV, the COI method relies solely on the macro-level COI-RoCoF metric, maintaining a relatively low total operating cost ($232,559) at the expense of nodal frequency security, which introduces significant operational risks. In contrast, the proposed method coordinates multiple inertia resources, increasing the total system operating cost by approximately 11.20% (from $232,559 to $258,617), but effectively ensures the nodal frequency security of the low-inertia system.

TABLE IV

COMPARISON OF ROCOF AND COSTS UNDER LOW INERTIA SCENARIOS

| Method | COI-RoCoF under most severe fault (Hz/s) | Maximum nodal RoCoF under most severe fault (Hz/s) | Generation cost ($) | Inertia dispatch cost ($) | Total cost ($) | Total inertia revenue ($) |
|---|---|---|---|---|---|---|
| COI | 1.0000 | 2.3500 | 231542 | 1017 | 232559 | 1017 |
| Proposed | 0.4989 | 1.0000 | 244289 | 14328 | 258617 | 44186 |

3) Distribution Pattern Analysis of Nodal Inertia Prices Under Extreme Fault Disturbances

Both reducing generator active power output and deploying VIR can improve nodal RoCoF security. To decouple these effects and further examine the influence of network topology on the spatial distribution of nodal inertia prices, this case study constructs multiple operating conditions in Scenario 2 by setting each generating unit, in turn, to its maximum active power output. Under each operating condition, a full set of unit-trip contingencies is considered to emulate system frequency response and inertia price formation under severe fault disturbances. Fig. 5 presents the inertia clearing prices at each VIR node under these contingencies. The results show that when G501, G502, or G503 in Area 5 is involved in the analyzed contingencies, the inertia clearing prices at VIR nodes across the system generally remain high. This is mainly because Area 5 has a relatively low inertia level. When a fault-induced trip occurs at a unit in this area, the system cannot rely on the remaining local inertia to effectively restrain the ensuing frequency transient, and therefore requires inertia support from remote nodes through tie-lines. However, because of the electrical impedance along the transmission paths, the supporting effect of remote inertia is significantly weakened during cross-area transfer. To satisfy the RoCoF security constraints for the entire system, especially for nodes in this weak area, the optimization model is forced to extensively dispatch higher-priced VIR resources across the entire grid. This inefficient cross-area support leads to tighter system-wide frequency security constraints, which subsequently drives a significant increase in the Lagrange multipliers globally, ultimately resulting in generally higher clearing prices.

From the spatial distribution of nodal inertia prices, it can be seen that these prices not only reflect the regional inertia scarcity under specific operating conditions but also quantify the marginal contribution of inertia resources at each node to maintaining nodal frequency security. From a planning and investment perspective, nodes that consistently exhibit high inertia prices under various extreme disturbance scenarios (such as Node 203, Node 102, etc.) possess the most significant nodal marginal value. Therefore, in the planning stage of future power systems with high shares of renewable energy, priority should be given to deploying VIR devices or new energy storage systems at such critical nodes, in order to achieve synergistic integration of enhanced system frequency security and optimal investment returns.

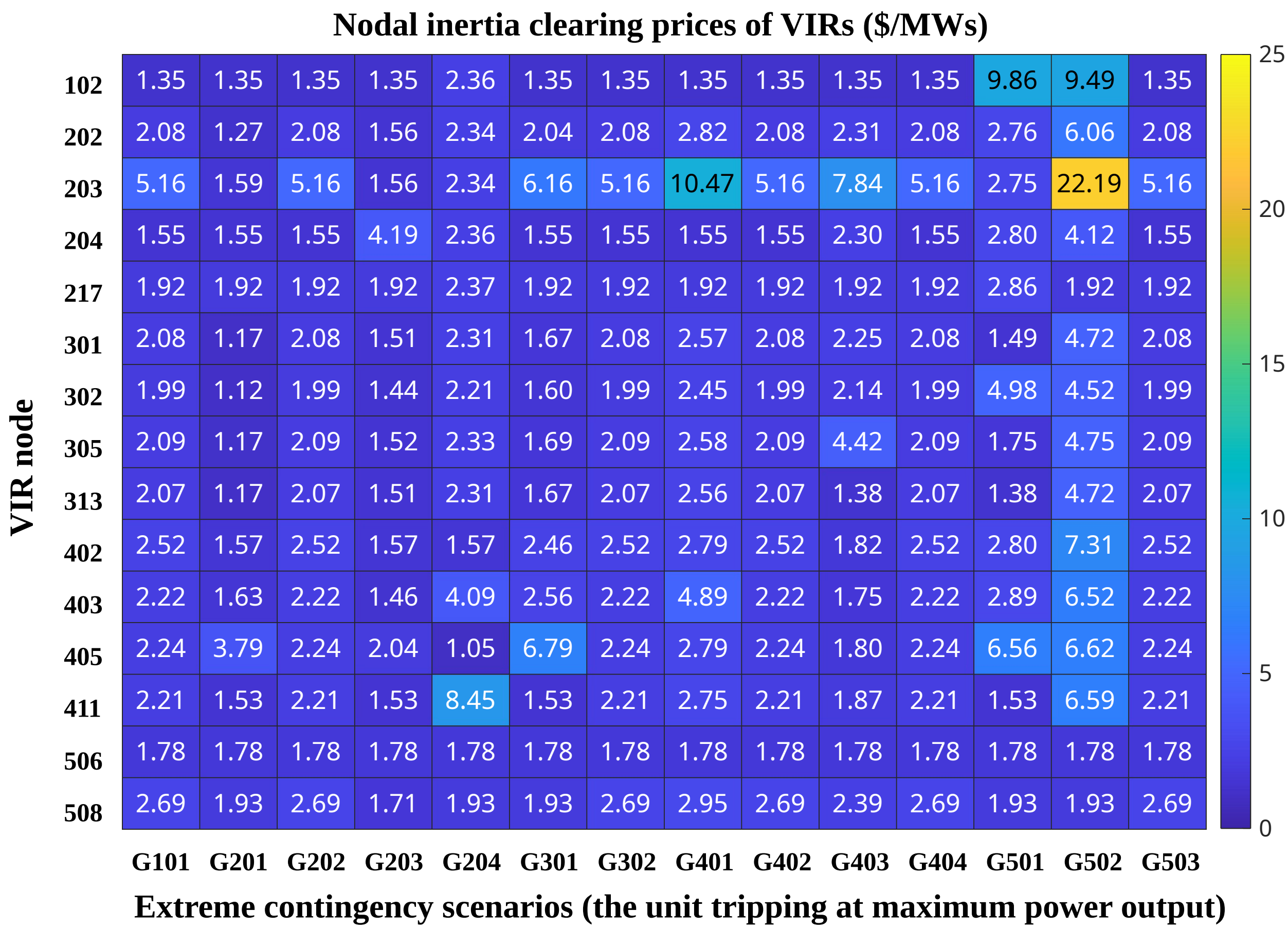

Fig. 5. Nodal inertia clearing prices of VIRs under different extreme contingency scenarios.

## V. Conclusion

This paper investigates the roles of synchronous and virtual inertia resources in the first stage following a disturbance by pricing their nodal security value and optimizing their allocation across nodes. A nodal RoCoF model is introduced in place of the traditional COI model, providing a more spatially resolved and physically grounded representation of post-disturbance frequency dynamics. The resulting nodal inertia prices better reflect the locational value of inertia and improve the consistency between market signals and actual frequency-security needs. Case studies under both high- and low-inertia scenarios show that, compared with the COI-based method, the proposed framework can more accurately identify inertia-weak nodes and effectively maintain nodal frequency security through the coordinated dispatch of synchronous and virtual inertia resources, with only a modest increase in operating cost. In addition, the spatial distribution of nodal inertia prices is closely linked to network topology and the placement of virtual inertia resources, suggesting that the proposed method can provide informative price signals for both short-term operation and the long-term siting and investment of virtual inertia resources.